\documentclass[reprint,prd,twocolumn,superscriptaddress,showpacs,floatfix]{revtex4}

\usepackage{amssymb}
\usepackage{graphicx}
\usepackage{epsfig}
\usepackage{multirow}
\newcommand{\ba}{\begin{eqnarray}}
\newcommand{\ea}{\end{eqnarray}}
\newcommand{\be}{\begin{equation}}
\newcommand{\ee}{\end{equation}}
\newcommand{\beq}{\begin{equation}} 
\newcommand{\eeq}{\end{equation}}   
\newcommand{\bea}{\begin{eqnarray}} 
\newcommand{\eea}{\end{eqnarray}}
\def\Li2{\hbox{Li}_2}

\begin{document}


\title{$\chi_{c1}$ and $\chi_{c2}$ production at $e^+e^-$ colliders. 
  }

\thanks{Work 
supported in part by
the Polish National Science Centre, grant number DEC-2012/07/B/ST2/03867 and
German Research Foundation DFG under
Contract No. Collaborative Research Center CRC-1044. }


\author{Henryk Czy\.z}
\affiliation{Institute of Physics, University of Silesia,
PL-40007 Katowice, Poland.}
\author{Johann H. K\"uhn}
\affiliation{Institut f\"ur Theoretische Teilchenphysik,
Karlsruhe Institute of Technology, D-76128 Karlsruhe, Germany.}
\author{Szymon Tracz}
\affiliation{Institute of Physics, University of Silesia,
PL-40007 Katowice, Poland.}


\date{\today}

\begin{abstract}
Direct, resonant production of the charmonium states $\chi_{c1}$ and
 $\chi_{c2}$ in electron-positron annihilation is investigated. Depending on
 details of the model, a sizeable variation of the prediction for the
 production cross section is anticipated. It is demonstrated that resonant
 production could be observed under favorable circumstances.
 
\end{abstract}

\pacs{13.66.Bc, 13.40.Gp }

\maketitle

\newcommand{\Eq}[1]{Eq.(\ref{#1})} 

\section{\label{sec1}Introduction}
The exclusive production of narrow resonances in electron-positron annihilation has been up to now observed for states with the quantum numbers of the virtual photon, $J^{PC}=1^{--}$, only. In principle axial vector resonances with $J^{PC}=1^{++}$ can be produced directly through two distinctly different mechanisms: either electromagnetically through two virtual photons or through the neutral current. The tensor state with $J^{PC}=2^{++}$, in contrast, can be produced through the electromagnetic process only. In practice, however, the rates are tiny at low energies and up to now only resonant production of hadrons with $J^{PC}=1^{--}$ has been observed experimentally. Nevertheless, already quite early the production of $1^{++}$ and $2^{++}$ states has been suggested, either through the neutral current \cite{Kaplan:1978wu} or through two virtual photons \cite{Kaplan:1978wu,Kuhn:1979bb}, with emphasis on charmonium resonances. In view of the small resonance enhancement, which is below or at most at the percent level, no experimental attempt has been made up to now to verify the predictions. However, with the advent of $e^+ e^-$ colliders with extremely high luminosity like BESIII, the picture has changed and this possibility has gained renewed interest \cite{Kivel:2015iea,Denig:2014fha,Yang:2012gk}. It now seems that resonant production of $\chi_{c1}$ and $\chi_{c2}$ might eventually be accessible by experiments. The signal could be observed either in a resonant excess of the hadronic cross section $e^+ e^- \to \chi_{c_J} \to$ hadrons or, alternatively, of the cross section $e^+ e^-\to \chi_{c_J}\to J/\psi + \gamma$ with subsequent decay $J/\psi \to \mu^+ \mu^-$. Note, that the interference with the continuum cross section $e^+ e^-\to J/\psi +\gamma$, which is the result of obvious radiative corrections, might play an important role in this connection.

\begin{figure}
\begin{center}
\includegraphics[width=9.cm]{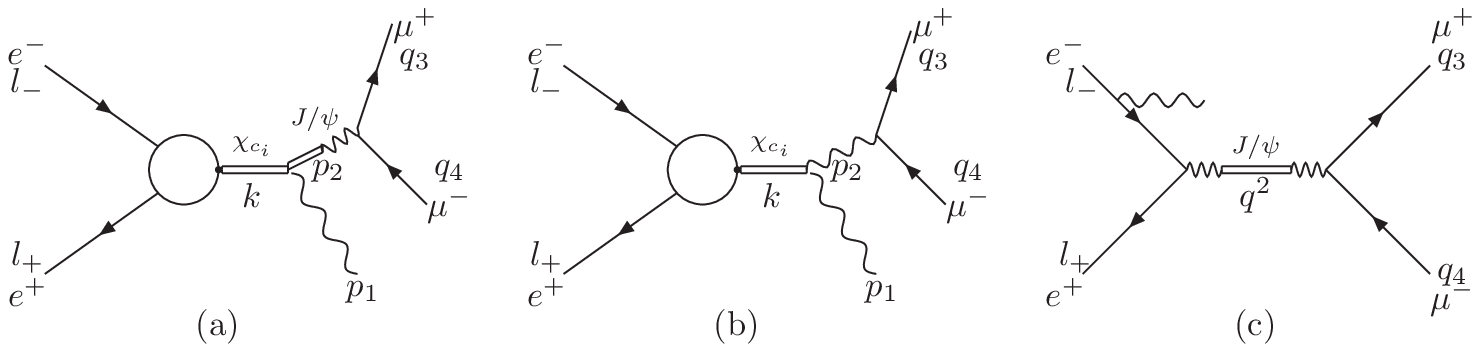}
\caption{Diagrams for the cross section for the process $e^+ e^-\to \chi_{c_J}\to \gamma J/\psi (\to \mu^+ \mu^-)$.
\label{cross}
}
\end{center}
\end{figure}

 It is the purpose of this paper to investigate these possibilities in detail. We will first evaluate the resonant electromagnetic cross section both for the $J^{PC}=1^{++}$ and the $J^{PC}=2^{++}$ state, including the influence of interference with continuum reaction (Figure \ref{cross}), recalling and extending earlier results \cite{Kaplan:1978wu,Kuhn:1979bb,Kivel:2015iea,Denig:2014fha,Yang:2012gk}. Two different final states will be considered: the hadronic cross section from the resonant reaction $e^+ e^- \to \chi_{c_J} \to$ hadrons, and the lepton plus photon state $e^+ e^-\to \chi_{c_J}\to \gamma J/\psi (\to \mu^+ \mu^-)$ together with its interference with the continuum $e^+ e^-\to \gamma J/\psi (\to \mu^+ \mu^-)$. Of course the $e^+ e^-$ energy has to be chosen equal to the mass of $\chi_{c_1}$ or $\chi_{c_2}$ and the photon energy has to be chosen in the
 proper kinematic region.

\section{\label{sec2}Resonant $\chi_{c_J}$ production}
\subsection{Short distance approximation}
Let us in a first step recall the results from \cite{Kuhn:1979bb,Yang:2012gk} on resonant $\chi_{c_J}$ production, using as a rough approximation the short distance expansion as discussed in \cite{Kuhn:1979bb}. The coupling to two virtual photons is given by 
\begin{eqnarray}
A_0^{\alpha\beta}(p_1,p_2)\epsilon^1_{\alpha}\epsilon^2_{\beta}&=&\sqrt{\frac{1}{6}}c\frac{2}{M_{\chi_{c_0}}}\{[(\epsilon_1\epsilon_2)(p_1p_2) \nonumber \\&&-(\epsilon_1p_2)(\epsilon_2p_1)][M_{\chi_{c_0}}^2+(p_1p_2)] \nonumber \\&&+(\epsilon_1p_2)(\epsilon_2p_2)p_1^2 +(\epsilon_1p_1)(\epsilon_2p_1)p_2^2\nonumber \\
&&  -(\epsilon_1\epsilon_2)p_1^2p_2^2 -(\epsilon_1p_1)(\epsilon_2p_2)p_1p_2\nonumber \},\\
 \label{amp0}
\end{eqnarray}
\begin{eqnarray}
A_1^{\alpha\beta}(p_1,p_2,\epsilon)\epsilon^1_{\alpha}\epsilon^2_{\beta}&=&ic\{ p_1^2(\epsilon,\epsilon_1,\epsilon_2,p_2)+p_2^2(\epsilon,\epsilon_2,\epsilon_1,p_1) \nonumber \\
&&\kern-32pt+\epsilon_1p_1(\epsilon,\epsilon_2,p_1,p_2)+\epsilon_2p_2(\epsilon,\epsilon_1,p_2,p_1)\}\nonumber,\\
 \label{amp1}
\end{eqnarray}
\begin{eqnarray}
A_2^{\alpha\beta}(p_1,p_2,\epsilon)\epsilon^1_{\alpha}\epsilon^2_{\beta}&=&\sqrt{2}cM_{\chi_{c_2}}\{(p_1p_2)\epsilon^1_{\mu} \epsilon^2_{\nu} \nonumber \\&+&p_{1\mu}p_{2\nu}(\epsilon_1\epsilon_2) \nonumber\\
&-& p_{1\mu}\epsilon_{\nu}^2(\epsilon_1p_2))-p_{2\mu}\epsilon_{\nu}^1(\epsilon_2p_1))\}\epsilon^{\mu\nu} \nonumber,\\
 \label{amp2} &&
\end{eqnarray}
where 
\begin{eqnarray}
c &\equiv&c((p_1+p_2)^2,p_1^2,p_2^2,m) \nonumber \\
&=&\frac{16\pi\alpha a}{\sqrt{m}}\frac{1}{((p_1-p_2)^2/4-m^2+i\epsilon)^2}, \nonumber\\
&& 
\label{eqcc}
\end{eqnarray}
with $m$ 
 the effective charm quark mass in $\chi_{c_i}$,
  $a=\sqrt{\frac{1}{4\pi}}3 Q^2\phi'(0)$, $\phi'(0)$ the derivative
 of the wave function at the origin
   and  $Q=2/3$ the charm quark
 electric charge. $p_1^2$, $p_2^2$, $\epsilon_1$, $\epsilon_2$ are the squares of the momenta and the polarization vectors of the photons and $\epsilon$ is the polarization vector in case of $\chi_{c_1}$ and the polarization tensor in case of $\chi_{c_2}$. We have checked that terms in the amplitudes, which are proportional to the binding energies and neglected in \cite{Kuhn:1979bb}, are breaking gauge invariance. Thus the results, Eqs.(\ref{amp0}-\ref{amp2}), do contain all the allowed binding energy corrections. Using this form of the photon resonance coupling, the amplitude for electron-positron annihilation is given by a loop integral and can be cast into the form:
\begin{eqnarray}
A(e^+ e^- \to^3P_J)&=&ie^2\int\frac{dp_1}{(2\pi)^4}\bar{v}(l_+)\gamma_{\nu}\not{h}\gamma_{\mu}u(l_-)\\
&& \frac{1}{h^2}\frac{1}{p_1^2}\frac{1}{p_2^2}A_J^{\mu\nu}(p_1,p_2,\epsilon),\nonumber
\end{eqnarray}
with $h=l_--p_1$.
Since we neglect the electron mass throughout, the amplitudes are given by
\begin{eqnarray}
A(e^+ e^- \to^3P_0)&=&0,\\
A(e^+ e^- \to^3P_1)&=&g_1\bar{v}\gamma_5/{\kern-5pt\epsilon}u,\\
A(e^+ e^- \to^3P_2)&=&g_2\bar{v}\gamma^{\mu}u\epsilon_{\mu\nu}(l_+^{\nu}-l_-^{\nu})/M_{\chi_{c_2}}.
\end{eqnarray}
As shown in \cite{Yang:2012gk} the mass corrections are completely negligible for electrons.
For the coefficients characterizing the amplitudes one finds \cite{Kuhn:1979bb}
\begin{eqnarray}
g_1&=&-\frac{\alpha^2\sqrt{2}}{M_{\chi_{c_1}}^{5/2}}32a\log\frac{2b_1}{M_{\chi_{c_1}}} \label{g1ap},\\
g_2&=&\frac{\alpha^2}{M_{\chi_{c_2}}^{5/2}}64a [\log\frac{2b_2}{M_{\chi_{c_2}}}+\frac{1}{3}(i\pi+\log{2}-1)]\label{g2ap}\nonumber,\\ &&
\end{eqnarray}
 with binding energy defined as $b_i = 2m-M_{\chi_{c_i}}$.
Notice that the definition of $a$ in \cite{Kuhn:1979bb} is different by a factor $\sqrt{3m}Q^2$ from the definition used here. 
The electronic widths are given by
\begin{equation}
\Gamma(^3P_1 \rightarrow e^+e^-)=\frac{1}{3}\frac{|g_1|^2}{4\pi}M_{\chi_{c_1}},
\label{gam1e}
\end{equation}
\begin{equation}
\Gamma(^3P_2 \rightarrow e^+ e^-)=\frac{1}{5}\frac{|g_2|^2}{8\pi}M_{\chi_{c_2}}.
\label{gam2e}
\end{equation}
Note that the result for $J=2$ differs from the one of \cite{Kuhn:1979bb} by a factor 2. Furthermore the factor $3Q_i^4$ has been taken into account in the definition of $a$.
The numerical results  are expected to depend significantly on the precise value of the charmed quark mass and the relative size of the absorptive part. For negative value of b the amplitudes develops a sizeable absorptive part which subsequently simulates the contribution from the intermediate state $J/\psi +\gamma$.
\subsection{Binding energy corrections}
In the next step we include binding energy corrections into the result. We thus include terms of order $1-x$ with $x=\frac{4m^2}{M_{\chi_{c_i}}^2}$. The decay rates are now given  by:
\begin{equation}
\Gamma(\chi_{c_1} \rightarrow e^+e^-)=\frac{1}{3}\frac{|g_{1_{\gamma\gamma}}|^2}{4\pi}M_{\chi_{c_1}},
\end{equation}
\begin{equation}
\Gamma(\chi_{c_2}\rightarrow e^+ e^-)=\frac{1}{5}\frac{|g_{2{\gamma\gamma}}|^2}{8\pi}M_{\chi_{c_2}}.
\end{equation}
with the coupling $g_{1_{\gamma\gamma}}$ and $g_{2_{\gamma\gamma}}$ given by
\bea
g_{1_{\gamma \gamma}}&=&  \frac{16\alpha^2a}{\sqrt{m}M_{\chi_{c_1}}^2}\Bigg[ 
  \log\left(\frac{x}{1+x}\right)
   \left(1-x\right)\nonumber\\
  &&-\left(\log\left(\frac{x}{1-x}\right)+i\pi\right)
   \left(1+x\right)\Bigg] ,
\eea

\bea
  g_{2_{\gamma \gamma}}&=&   \frac{32\sqrt{2}\alpha^2a}{3\sqrt{m}M_{\chi_{c_2}}^2} 
 \Bigg[\left( \frac{1+x}{2}+\frac{8}{(1+x)^2}\right)\log(1-x)\nonumber\\
 &&\kern-32pt+ \frac{3}{2}\left(1+x\right) \log(1+x)
 -2\left(1+x + \frac{2}{(1+x)^2}\right)\log(x)\nonumber\\
 &&\kern-32pt-\frac{8}{(1+x)^2}\log(2) -1 
  -\frac{i\pi}{2}\left( 1+x +\frac{8}{(1+x)^2}\right)
\Bigg]\nonumber.\\
 \eea
Of course, in the limit $ x\to 1$ the results from equations (\ref{g1ap}) and (\ref{g2ap}) are recovered.
Leading order approximation and exact results for positive and negative binding energy are given in Table \ref{npb_tab1}, where we have used a typical value of $0.1 GeV^{5}$ for $|\phi^`(0)|^2$. As one can see the binding
  energy corrections, which are usually neglected \cite{Kaplan:1978wu,Kuhn:1979bb,Kivel:2015iea,Denig:2014fha,Yang:2012gk},
are not negligible and can amount up to almost 50\% of the leading result.
  Moreover
they have opposite sign for $\chi_{c1}$ and $\chi_{c2}$ case leading to
 substantial difference between the $\chi_{c1}$ and $\chi_{c2}$
  widths.
\begin{table}
\begin{center}
\vskip0.3cm
\begin{tabular}{|c|c|c|}
\hline
&$\Gamma(\chi_{c_1}\rightarrow e^+ e^-)$ & $\Gamma(\chi_{c_2}\rightarrow e^+ e^-)$ \\
\hline
&\multicolumn{2}{|c|}{$b=0.5\  GeV$} \\
\hline
Leading term & 0.0226\ eV & 0.0243\ eV \\
exact result &0.0317\ eV & 0.0159\ eV\\
\hline
&\multicolumn{2}{|c|}{$b=-0.5\  GeV$}\\
\hline
Leading term &0.164\ eV & 0.0512\ eV \\
exact result &0.141\ eV & 0.0731\ eV\\
\hline
\end{tabular}
\caption{Electronic widths for $b=-0.5 GeV$ and $b=0.5 GeV$ }
\label{npb_tab1}
\end{center}
\end{table}

\subsection{Short and Long distance combined}
\begin{figure}
\begin{center}
\includegraphics[width=7.5cm]{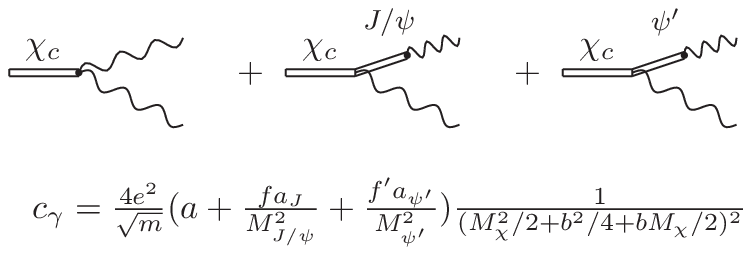}
\caption{Diagrams for decay widths $\Gamma(\chi_{c_{0,1,2}} \to \gamma \gamma )$.
\label{gg_4}
}
\end{center}
\end{figure}

\begin{figure}
\begin{center}
\includegraphics[width=8.cm]{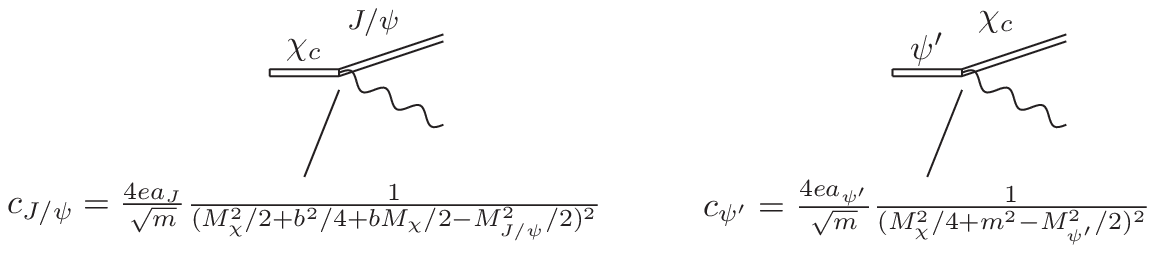}
\caption{Diagrams for decay widths $\Gamma(\chi_{c_{0,1,2}} \to \gamma J/\psi)$.
\label{jpsi_4}
}
\end{center}
\end{figure}

\begin{figure}
\begin{center}
\includegraphics[width=2.5cm]{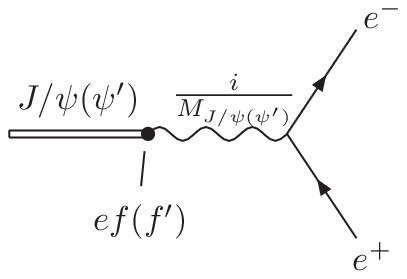}
\caption{Diagram for decay width  $\Gamma(J/\psi(\psi') \to e^+ e^- )$ .
\label{psiee}
}
\end{center}
\end{figure}

Although the model discussed in the previous section exhibits the correct leading logarithmic behavior of the photon-photon -$\chi_{c_i}$ coupling, the non-enhanced terms are of comparable size, potentially even larger than the formally dominant ones. For this reason we formulate an ansatz which gives the correct behavior for the coupling of $\chi_{c_2}$ to two photons and for the coupling of both $\chi_{c_1}$ and $\chi_{c_2}$ to $J/\psi \gamma$. We start from the following ansatz (see Figs. \ref{gg_4}, \ref{jpsi_4}):
\bea
A_{1\gamma \gamma}^{\alpha\beta}(p_1,p_2,\epsilon)\epsilon^1_{\alpha}\epsilon^2_{\beta}\bigg{|}_{p_1^2=p_2^2=0}\kern-32pt&=&0,
\eea
\bea
A_{1\gamma J/\psi}^{\alpha\beta}(p_1,p_2,\epsilon)\epsilon^1_{\alpha}\epsilon^2_{\beta}\bigg{|}_{p_1^2=0,\ p_2^2=M_{J/\psi}^2}\kern-32pt&=&ic_{J/\psi}^1\Big\{ p_2^2(\epsilon,\epsilon_2,\epsilon_1,p_1)\nonumber\\ &&+\epsilon_1p_1(\epsilon,\epsilon_2,p_1,p_2) \nonumber \\
&&+\epsilon_2p_2(\epsilon,\epsilon_1,p_2,p_1)\Big\}\nonumber,\\
&&
\label{chi1Jg}
\eea
\bea
A_{2\gamma \gamma}^{\alpha\beta}(p_1,p_2,\epsilon)\epsilon^1_{\alpha}\epsilon^2_{\beta}\bigg{|}_{p_1^2=p_2^2=0}\kern-32pt&=&\sqrt{2}c_{\gamma}^2M_{\chi_{c_2}}\Big\{(p_1p_2)\epsilon^1_{\mu} \epsilon^2_{\nu} \nonumber \\&+&p_{1\mu}p_{2\nu}(\epsilon_1\epsilon_2) \nonumber\\
&-& p_{1\mu}\epsilon_{\nu}^2(\epsilon_1p_2))-p_{2\mu}\epsilon_{\nu}^1(\epsilon_2p_1))\Big\}\epsilon^{\mu\nu} \nonumber,\\
&&
\eea
\bea
A_{2\gamma J/\psi}^{\alpha\beta}(p_1,p_2,\epsilon)\epsilon^1_{\alpha}\epsilon^2_{\beta}\bigg{|}_{p_1^2=0,\ p_2^2=M_{J/\psi}^2}\kern-42pt&=&\sqrt{2}c_{J/\psi}^2M_{\chi_{c_2}}\Big\{(p_1p_2)\epsilon^1_{\mu} \epsilon^2_{\nu}\nonumber \\&+&p_{1\mu}p_{2\nu}(\epsilon_1\epsilon_2) \nonumber\\
&&\kern-32pt- p_{1\mu}\epsilon_{\nu}^2(\epsilon_1p_2))-p_{2\mu}\epsilon_{\nu}^1(\epsilon_2p_1))\Big\}\epsilon^{\mu\nu} \nonumber,\\
&&
\label{chi2Jg}
\eea
where, in the case of the amplitudes $A_{i\gamma\gamma}$, $p_1$ and $p_2$ are the momenta of photons, $\epsilon_1$ and $\epsilon_2$ are their polarization vectors. In the case of the amplitudes $A_{i\gamma J/\psi }$, $p_1$ is the photon momentum, $\epsilon_1$ its polarization vector, $p_2$ is the $J/\psi$ momentum and $\epsilon_2$ its polarization vector. The function $c_{\gamma}$ is the $\chi_{c_i} - \gamma \gamma$ form factor, whereas $c_{J/\psi}$ is the $\chi_{c_i} - \gamma J/\psi$ form factor. These form factors have the following forms:
\begin{eqnarray}
 &&\kern -30pt c_{\gamma}^i\equiv (1+\frac{f\cdot a_J}{aM_{J/\psi}^2}+\frac{f'\cdot a_{\psi'}}{aM_{\psi'}^2})c(M_{\chi_{c_i}}^2,0,0,m) = \nonumber \\
  &&\kern -30pt   \frac{16\pi\alpha}{\sqrt{ m}}( a+\frac{f\cdot a_J}{M^2_{J/\psi}}+\frac{f'\cdot a_{\psi'}}{M_{\psi'}^2})\nonumber\\
  &&\frac{1}{\left(M^2_{\chi_{c_i}}/2+b^2_i/4+b_iM_{\chi_{c_i}}/2\right)^2},
 \label{cccg}
\end{eqnarray}

\begin{eqnarray}
 &&\kern -30pt c_{J/\psi}^i\equiv\frac{a_J}{ae} c(M_{\chi_{c_i}}^2,0,M_{J/\psi}^2,m)  = \nonumber \\
  &&\kern -30pt   \frac{4ea_J}{\sqrt{ m}}
  \frac{1}{\left(M^2_{\chi_{c_i}}/2+b^2_i/4+b_iM_{\chi_{c_i}}/2-M_{J/\psi}^2/2\right)^2}. \nonumber \\
&& 
 \label{cccpsi}
\end{eqnarray} 
The couplings $a_J$ and $a_{\psi'}$ in our model are free parameters. 
They  can be related,
 similarly to \cite{Kivel:2015iea}, to overlap of the radial wave functions
 calculated in a framework of a potential model 
   \cite{Eichten:1978tg}. Yet, as any potential model has free parameters
 to be fitted, we prefer to extract $a_J$ and $a_{\psi'}$ 
 directly from experimental data.  
As it was shown in \cite{Kivel:2015iea}, the $\psi'$ contributions to the $\chi_c$ electronic widths are important. We model the $\psi'\to \chi_c \gamma$ amplitudes in an analogous way to Eq.(\ref{chi1Jg}) and Eq.(\ref{chi2Jg})
\bea
A_{\psi'1\gamma}^{\alpha\beta}(p_1,p_2,\epsilon)\epsilon^1_{\alpha}\epsilon^2_{\beta}\bigg{|}_{p_1^2=0,\ p_2^2=M_{{\psi'}}^2}\kern-32pt&=&ic_{\psi'}^1\Big\{ p_2^2(\epsilon,\epsilon_2,\epsilon_1,p_1)\nonumber\\ &&+\epsilon_1p_1(\epsilon,\epsilon_2,p_1,p_2) \nonumber \\
&&+\epsilon_2p_2(\epsilon,\epsilon_1,p_2,p_1)\Big\}\nonumber,\\
&&
\label{psipchi1g}
\eea
\bea
A_{\psi'2\gamma}^{\alpha\beta}(p_1,p_2,\epsilon)\epsilon^1_{\alpha}\epsilon^2_{\beta}\bigg{|}_{p_1^2=0,\ p_2^2=M_{\psi'}^2}\kern-42pt&=&\sqrt{2}c_{\psi'}^2M_{\chi_2}\Big\{(p_1p_2)\epsilon^1_{\mu} \epsilon^2_{\nu}\nonumber \\&+&p_{1\mu}p_{2\nu}(\epsilon_1\epsilon_2) \nonumber\\
&&\kern-32pt- p_{1\mu}\epsilon_{\nu}^2(\epsilon_1p_2))-p_{2\mu}\epsilon_{\nu}^1(\epsilon_2p_1))\Big\}\epsilon^{\mu\nu} \nonumber,\\
&&
\label{psipchi2g}
\eea
where
\begin{eqnarray}
 &&\kern -30pt c_{\psi'}^i =  \frac{4ea_{\psi'}}{\sqrt{ m}}
  \frac{1}{\left(M^2_{\chi_{c_i}}/4+m^2-M_{\psi'}^2/2\right)^2}, \nonumber \\
&& 
 \label{cccpsip}
\end{eqnarray}
where $M_{\psi'}$ is the $\psi'(\equiv \psi(2S))$ mass. 
With this ansatz one obtains

\begin{eqnarray}
&& \kern -30pt\Gamma(\chi_{c_1} \rightarrow J/\psi \gamma)= \nonumber \\
 &&\frac{1}{96\pi}|c_{J/\psi}^1|^2M_{J/\Psi}^2M_{\chi_{c_1}}^3(1+x_1)(1-x_1)^3, \label{w1}\\
&& \kern -30pt\Gamma(\chi_{c_2} \rightarrow \gamma \gamma)=\frac{1}{160\pi}|c_{\gamma}^2|^2M_{\chi_{c_2}}^5,\label{w2} \\
&&\kern -30pt 
 \Gamma(\chi_{c_2} \rightarrow J/\psi \gamma) = \nonumber \\
&&\frac{1}{80\pi}|c_{J/\psi}^2|^2M_{\chi_{c_2}}^5(1-x_2)^3(1+x_2/2+x_2^2/6) \label{w3},\\
&& \kern -30pt\Gamma(\psi' \rightarrow \chi_{c_1} \gamma)= \nonumber \\
 &&\frac{1}{96\pi}|c_{\psi'}^1|^2M_{\psi'}^5(1+\bar x_1)(1-\bar x_1)^3/\bar x_1, \label{wp1}
\\
&&\kern -30pt 
 \Gamma(\psi' \rightarrow\chi_{c_2} \gamma) = \nonumber \\
&&\frac{1}{288\pi}|c_{\psi'}^2|^2M_{\psi'}^5(1-\bar x_2)^3(1+3\bar x_2+6\bar x_2^2)/\bar x_2 ,\label{wp2}
\end{eqnarray}
where $x_i=M_{J/\Psi}^2/M_{\chi_{c_i}}^2$, $\bar x_i=M_{\chi_{c_i}}^2/M_{\psi'}^2$ and $c_{\gamma}^i,\ c_{J/\psi}^i$, $c_{\psi'}^i$ are defined in Eq. (\ref{cccg}), (\ref{cccpsi}) and (\ref{cccpsip}). The parameter $a$ has been defined after Eq.(\ref{eqcc}). 
The constants $f$ and $f'$ have been extracted from the electronic widths
 of  $J/\psi$  and $\psi'$ calculated according to the diagram from Fig. \ref{psiee} and have the following form:
\begin{equation}
f=\sqrt{\frac{3\Gamma_{J/\psi\rightarrow e^+ e^-}M_{J/\psi}^3}{4\pi\alpha^2}}
; \ f'=\sqrt{\frac{3\Gamma_{\psi'\rightarrow e^+ e^-}M_{\psi'}^3}{4\pi\alpha^2}}.
\end{equation}

The $R_{peak}$ value at the peak of the cross section is given by \cite{Buchmuller:1987nq} 
\begin{equation}
R_{peak}=\frac{\sigma_{res}^{(0)}}{\sigma_{pt}}=\frac{\Gamma_{ee}}{\Delta}\frac{9}{4\alpha^2}\sqrt{2M}\frac{\Gamma_{had}}{\Gamma_{tot}}N_Z
\end{equation}
where $\Gamma_{ee}, \Gamma_{had}$ and $\Gamma_{tot}$ denote the width of the resonance into $e^+e^-$, into hadrons and the total width, respectively. $\Delta$ stands for the machine energy resolution and $N_Z$ is slightly model dependent factor around 0.7. Taking for illustration values for $\Gamma_{ee}$ between 0.1 eV and 0.5 eV, $\Gamma_{had}/\Gamma_{tot}=0.66$ and $\Delta=4 MeV$, one finds $R_{peak}$ between 2.15$\cdot 10^{-3}$ and 1.075$\cdot 10^{-2}$. 

Alternatively, one may focus on the decay channel $e^+ e^-\to \chi_{c_i}\to \gamma J/\psi (\to \mu^+ \mu^-)$. For the $1^{++}$ state the prediction is also affected by the amplitude due to the neutral current \cite{Kaplan:1978wu,Kuhn:1979bb,Yang:2012gk} . To identify the interference term, the neutral current amplitude has to be decomposed into the form $(V_e+A_e)A_C$, and it is the interference between the $A_eA_C$ term from the neutral current and the dispersive part (real part) of the electromagnetic amplitude which affects the rate. Specifically one obtains:
\bea
\kern-24pt\Gamma(\chi_{c_1} \rightarrow e^+e^-)&=&\frac{M_{\chi_{c_1}}}{3\pi}\Bigg[\frac{|g_1|^2}{4}\nonumber \\
&&\kern-64pt+\frac{aG_F}{\sqrt{2m}Q^2}Re(g1)  \nonumber \\
&&\kern-64pt+ \frac{a^2G_F^2}{mQ^4}\Bigg(1-4\sin^2{\theta_W}+8\sin^4{\theta_W}\Bigg) \Bigg],
\label{zzzzz}
\eea
where $G_F$ is the Fermi constant and $\theta_W$ is the weak mixing angle. The function $g_1$ comes from performing loop integrals (see Appendix \ref{g1g2}).\\
\hspace{2cm}
The mass of the $c$ quark, the derivative of the wave function at the origin (in fact $a$) and the parameters $a_J$ and $a_{\psi'}$ have been extracted from the measured decay widths \cite{Agashe:2014kda} of $\chi_{c_{1,2}}$ to $\gamma \gamma$ and to $\gamma J/\psi$ and of $\psi'$ to $\chi_{c_{1,2}} \gamma$, using formulae (\ref{w1})-(\ref{wp2}). The fit of 4 parameters to 5 experimental values
 has given $\chi^2=0.16$. \\
The obtained parameters, the square of the derivative of the wave function $|\phi^{'}(0)|^2$, the effective c-quark mass, and the parameters $a_{J(\psi')}$ are presented in Table \ref{fit24} together with the calculated decay widths.
 There exists another set of parameters giving the same $\chi^2$, but in this
 fit $a_{\psi'}$ is positive. As one knows from potential models 
  \cite{Eichten:1978tg} the  $a_{\psi'}$ should be negative. This result
  was independent on the parameters of the potential used 
  in \cite{Eichten:1978tg} and thus we use this information to reject 
  the fit parameters with positive $a_{\psi'}$.

\begin{table}
\begin{center}
\vskip0.3cm
\begin{tabular}{|c|c|c|c|c|}
\hline
 a[GeV$^{5/2}$] & $|\phi^{'}(0)|^2$ [GeV$^{5}$] & $m$ [GeV]& $a_J$[GeV$^{5/2}$] & $a_{\psi}$ [GeV$^{5/2}$]\\
\hline
 0.0786  & 0.04 & 1.69 & 0.15 & -0.07 \\
\hline
\end{tabular}
\begin{tabular}{|c|c|c|}
\hline
 widths [MeV] & $\chi_{c1}$ & $\chi_{c2}$   \\
\hline
\hline
$\Gamma(\chi\rightarrow \gamma \gamma)_{th}$ & - &$5.288 \cdot10^{-4}$\\
$\Gamma(\chi\rightarrow J /\psi \gamma)_{th}$ & $2.803 \cdot 10^{-1}$ & $3.778\cdot 10^{-1}$\\
$\Gamma(\psi'\rightarrow \chi \gamma)_{th}$ & $ 2.856 \cdot 10^{-2}$ & $2.705\cdot 10^{-2}$\\
\hline
$\Gamma(\chi\rightarrow \gamma \gamma)_{exp}$ & - &$5.3(3) \cdot10^{-4}$\\
$\Gamma(\chi\rightarrow J /\psi \gamma)_{exp}$ & $2.8(2) \cdot 10^{-1}$ & $3.7(3)\cdot 10^{-1}$\\
$\Gamma(\psi'\rightarrow \chi \gamma)_{exp}$ & $2.8(1) \cdot 10^{-2}$ & $2.7(1)\cdot 10^{-2}$\\
\hline
\end{tabular}

\caption{Parameters and theoretical ($th$) (this paper), and experimental ($exp$) \cite{Agashe:2014kda} values of $\Gamma(\chi_{1,2}\rightarrow \gamma \gamma, \gamma J/\psi )$ and $\Gamma(\psi'\rightarrow \chi_{1,2}\gamma )$.}
\label{fit24}
\end{center}

\end{table}

\begin{figure}

\begin{center}

\includegraphics[width=6.cm]{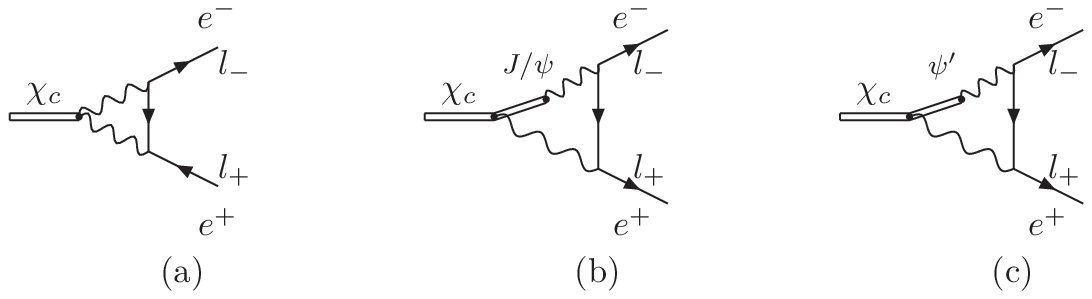}
\caption{Diagrams for decay widths $\Gamma(\chi_{c_{0,1,2}} \to e^+ e^- )$.
\label{el_jpsi4}
}
\end{center}
\end{figure}

The electronic widths have been calculated using the diagrams from Figure \ref{el_jpsi4}. For $\chi_{c_1}$ we have, in addition, also included the contribution coming from the neutral current Eq. (\ref{zzzzz}). The functions $g_i$, which come from performing loop integrals can be divided into three parts:
\begin{equation} 
g_i=g_{i_{\gamma\gamma}}+g_{i_{J/\psi\gamma}}+g_{i_{\psi'\gamma}},
\label{sum123}
\end{equation}
coming from Fig.\ref{el_jpsi4}a, Fig.\ref{el_jpsi4}b and Fig.\ref{el_jpsi4}c.
The formulae for these functions can be found in Appendix \ref{g1g2}.
In Table \ref{mod4_ee} we present the values of the electronic widths within the adopted model. The columns $\gamma\gamma$, $J/\psi\gamma$ and $\psi'\gamma$ give the individual rates from the contributions of the corresonding final states, the column QED gives the coherent sum.
 For $\chi_{c_1}$ we include the sum of electromagnetic and neutral current contribution ($QED+Z^0$). The obtained values of the electronic widths are much bigger than the ones obtained 
 within other models \cite{Kivel:2015iea,Denig:2014fha,Yang:2012gk} and
 definitively can be observed at BESIII scan experiments \cite{Denig:2014fha}
 (see also the next section).
 This is thus a matter of an experimental verification, which of the models
 is correct.
 

\begin{table}
\begin{center}
\vskip0.3cm
\begin{tabular}{|c|c|c|c|c|c|}
\hline
 & QED & $\gamma \gamma$ & $J/\psi\gamma$  & $\psi'\gamma$& QED+$Z^0$ \\ 
\hline
$\Gamma(\chi_{c_1}\to e^+ e^-)$ [eV]&0.43&0.10  &0.008 &0.094& 0.41 \\
$\Gamma(\chi_{c_2}\to e^+ e^-)$ [eV]&4.25&0.042  &1.41  & 0.45 & -\\
\hline
\end{tabular}
\caption{Electronic widths for $\chi_{c_1}$ and $\chi_{c_2}$.
 QED means the sum of $\gamma \gamma$, $J/\psi\gamma$  and $\psi'\gamma$
 contributions. See text for details.}
\label{mod4_ee}
\end{center}
\end{table}

\section{\label{sec3} The process $e^+ e^-\to \chi_{c_i}\to \gamma J/\psi (\to \mu^+ \mu^-)$}
With the couplings extracted as described above one can predict the $\chi_{c_1}$ and $\chi_{c_2}$
 production cross sections in $e^+e^-$ annihilation. As these states are not
 stable one can observe only their decay products and an easy to identify
 final state has to be chosen. An obvious choice is the reaction
  $e^+e^-\to \chi_c \to \gamma J/\psi (\to \mu^+\mu^-)$. The Feynman
  diagram describing this process is given in Fig.\ref{cross}a. In Fig.\ref{cross}b we present the diagram for the similar process, where $J/\psi$ is substituted by $\gamma$. The same
 final state is produced also in the ISR process (Fig. \ref{cross}c) and
  the amplitudes interfere.

 Within the adopted model the $\chi_{c_i}$ production amplitudes read

  \bea
\mathcal{M}_0&=&0, \\
\mathcal{M}_1&=&\Bigg\{g_1\bar{v}(l_+)\gamma_5 \gamma^{\mu}u(l_-) \nonumber \\
 && + \frac{2aG_FM_{\chi_{c_1}}^2}{h}\bar{v}(l_+)\Big((1+2m/M_{\chi_{c_1}})\gamma_5 \gamma^{\mu}\nonumber \\
&& +(1-4\sin^2{\theta_W}+2m/M_{\chi_{c_1}}\nonumber \\
&& -8m/M_{\chi_{c_1}}\sin^2{\theta_W})\gamma^{\mu}\Big)u(l_-)\Bigg\}  \nonumber \\
&&  \Pi_{\mu\nu}^{\chi_{c1}}(k) \ A_1^{\nu\beta} \ \Pi^{{J/\psi}}_{\beta\delta}(p_2)
 \ e \ \bar{u}(q_3)\gamma^{\delta}v(q_4)\nonumber \\
&& \\
\mathcal{M}_2&=&g_2\bar{v}(l_+)\gamma^{\mu}u(l_-)(l_+^{\nu}-l_-^{\nu})/M_{\chi_{c_2}}
  \nonumber \\
  && \kern -30pt \Pi_{\mu\nu\alpha\beta}^{\chi_{c2}}(k) \ A_2^{\alpha\beta\gamma}
 \ \Pi^{{J/\psi}}_{\gamma\delta}(p_2) \ e \ \bar{u}(q_3)\gamma^{\delta}v(q_4), 
\label{c1ampl}
\eea
where $h=2\sqrt{2}(M_{\chi_{c_1}}/2+m)M_{\chi_{c_1}}\sqrt{m}Q^2$. The amplitudes $A_i^{\nu\beta}$ can be found in Appendix B of 
 \cite{Kuhn:1979bb},
\bea
A_1^{\nu\beta}&=&-i\frac{1}{2}c({I_1^1}^{\nu\beta}+{I_2^1}^{\nu\beta}),\\
A_2^{\alpha\beta\gamma}&=&-c\sqrt{2}M_{\chi_{c_2}}I_2^{2\alpha\beta\gamma},
\label{cc}
\eea
and coincide with Eq.(\ref{amp1}) and Eq.(\ref{amp2}), with $c$ given
 in Eq.(\ref{cccpsi}).
Here the contributions $I_1^1$, $I_2^1$  and $I_2^{2}$ are given by:
\begin{eqnarray}
{I_1^1}^{\nu\beta}&=& \epsilon^{\bar{\mu}\bar{\nu}\beta\nu}F_{\bar{\mu}\bar{\nu}}^1p^2_{\bar{\gamma}}{p^2}^{\bar{\gamma}}-\epsilon^{\bar{\mu}\bar{\nu}\bar{\alpha}\nu}F_{\bar{\mu}\bar{\nu}}^1p^2_{\bar{\alpha}}{p^2}^{\beta},\\
{I_2^1}^{\nu\beta}&=&0,\\
I_2^{2\alpha\beta\gamma}&=& F^{1\alpha\delta}\left(g^{\beta\gamma}p^{2}_{\delta}-g_{\delta}^{\gamma}p^{2\beta}\right),
\end{eqnarray}
where
\begin{equation}
F_{\mu\nu}^1=\epsilon_{\mu}^1p_{\nu}^1-\epsilon_{\nu}^1p_{\mu}^1.
\label{Fuv}
\end{equation}
The ${I_2^1}^{\nu\beta}$ vanishes for one real photon in the vertex.
The coupling   of $J/\psi$ to muons and the $J/\psi$ propagator collected in $\Pi^{{J/\psi}}$, are given by,
\begin{equation}
\Pi^{{J/\psi}}_{\beta\delta}(p)=\sqrt{\frac{3\Gamma_{J/\psi\rightarrow e^+e^-}}{\alpha\sqrt{p_2^2}}}
 \frac{g_{\beta\delta}-p_\beta p_\delta/M_{J/\psi}^2}
 {p_2^2-M_{J/\psi}^2+iM_{J/\psi}\Gamma_{J/\psi}},
\end{equation}
while the $\chi_{c_1}$ propagator $\Pi^{\chi_{c1}}$ has the following form:
\begin{equation}
\Pi_{\mu\nu}^{\chi_{c1}}(k)=\frac{g_{\mu\nu}-k_\mu k_\nu/M_{\chi_{c1}}^2}{k^2-M^2_{\chi_{c_1}}+i\Gamma_{\chi_{c_1}} M_{\chi_{c_1}}}, 
\end{equation}
where $k$ is the four-momentum of the $\chi_{c_1}$, $M_{\chi_{c_1}}$ and
 $\Gamma_{\chi_{c_1}}$  are its mass
  and its decay width respectively.
The $\chi_{c_2}$ propagator $\Pi^{\chi_{c2}}$ has the following form:
\begin{equation}
\Pi^{\chi_{c2}}_{\mu\nu\alpha\beta}(k)=\frac{B_{\mu\nu\alpha\beta}}{k^2-M^2_{\chi_{c_2}}+i\Gamma_{\chi_{c_2}} M_{\chi_{c_2}}},
\end{equation}
where we use similar notation as for $\chi_{c_1}$. The tensor $B_{\mu\nu\alpha\beta}$ is given by the following formula:
\begin{equation}
B_{\mu\nu\alpha\beta}=\frac{1}{2}(P_{\mu\alpha}P_{\nu\beta}+P_{\mu\beta}P_{\nu\alpha})-\frac{1}{3}P_{\mu\nu}P_{\alpha\beta}),
\end{equation}
where $P_{\mu\nu}=-g_{\mu\nu}+k_{\mu}k_{\nu}/M_{\chi_{c_2}}$.
The form factor $c$ is given in Eq.(\ref{eqcc}).
\section{\label{sec5} Implementation into the PHOKHARA generator: 
 tests and results}

The amplitudes described in the previous section were implemented into
 the PHOKHARA event generator and will appear at the web page
 {\it (http://ific.uv.es/$\sim$rodrigo/phokhara/)} as release 9.2.
 The radiative return amplitude was already implemented in the version
 7.0 \cite{Czyz:2010hj,Czyz:2009vj}. The implementation of the
 other amplitudes was tested by constructing two independent codes:
 one using a trace method to sum over polarizations of initial and 
 final particles, the second one using the helicity amplitude method
 with a basis chosen as in \cite{Rodrigo:2001kf}. 
 Excellent agreement of relative
 accuracy about $10^{-15}$, was found except in the region
 where the amplitudes have zeros, but even for these negligible contributions
 several digits of the results agree. 

 Another test consisted of a comparison of the  integrated cross section
 obtained by the PHOKHARA generator and an analytic form 
 ($\sigma_{1,2}$) obtained 
 with the amplitude  from figure \ref{cross}a 
 for the scattering energy $\sqrt{s}=M_{\chi_{c_{1,2}}}$  in the narrow 
 width approximation given below
\bea
\sigma_1&=&\frac{12\pi}{s} Br(\chi_{c_1}\rightarrow e^+ e^-)\nonumber \\
                       &&Br(\chi_{c_1}\rightarrow J/\psi \gamma)
             Br(J/\psi \rightarrow \mu^+\mu^-), \\
\sigma_2&=&\frac{20\pi}{s} Br(\chi_{c_2}\rightarrow e^+ e^-)\nonumber \\
                       &&Br(\chi_{c_2}\rightarrow J/\psi \gamma)
        Br(J/\psi \rightarrow \mu^+\mu^-),
\label{sigma}
\eea
with the partial widths given in 
 Eqs.(\ref{w1},\ref{w3},\ref{gam1e},\ref{gam2e}) and 
\bea
&&\Gamma_{J/\psi \rightarrow \mu^+\mu^-}=\nonumber \\
&&(1+2\frac{m^2_{\mu}}{M_{J/\psi}^2})\sqrt{1-4m_{\mu}^2/M_{J/\psi}^2}.
 \ \Gamma_{J/\psi \rightarrow e^+ e^-}\label{jpsiwidth}.
\eea
The total widths are taken from \cite{Agashe:2014kda}.
 As all the widths here 
 are narrow, the approximation works well. The relative difference between
 the generator results and the analytic formulae Eq.(\ref{sigma}) are
 $ 3.4 \%$  for $\chi_{c_1}$ and  $1.3\%$ for $\chi_{c_2}$.

\begin{figure}
\begin{center}
\includegraphics[width=9.cm]{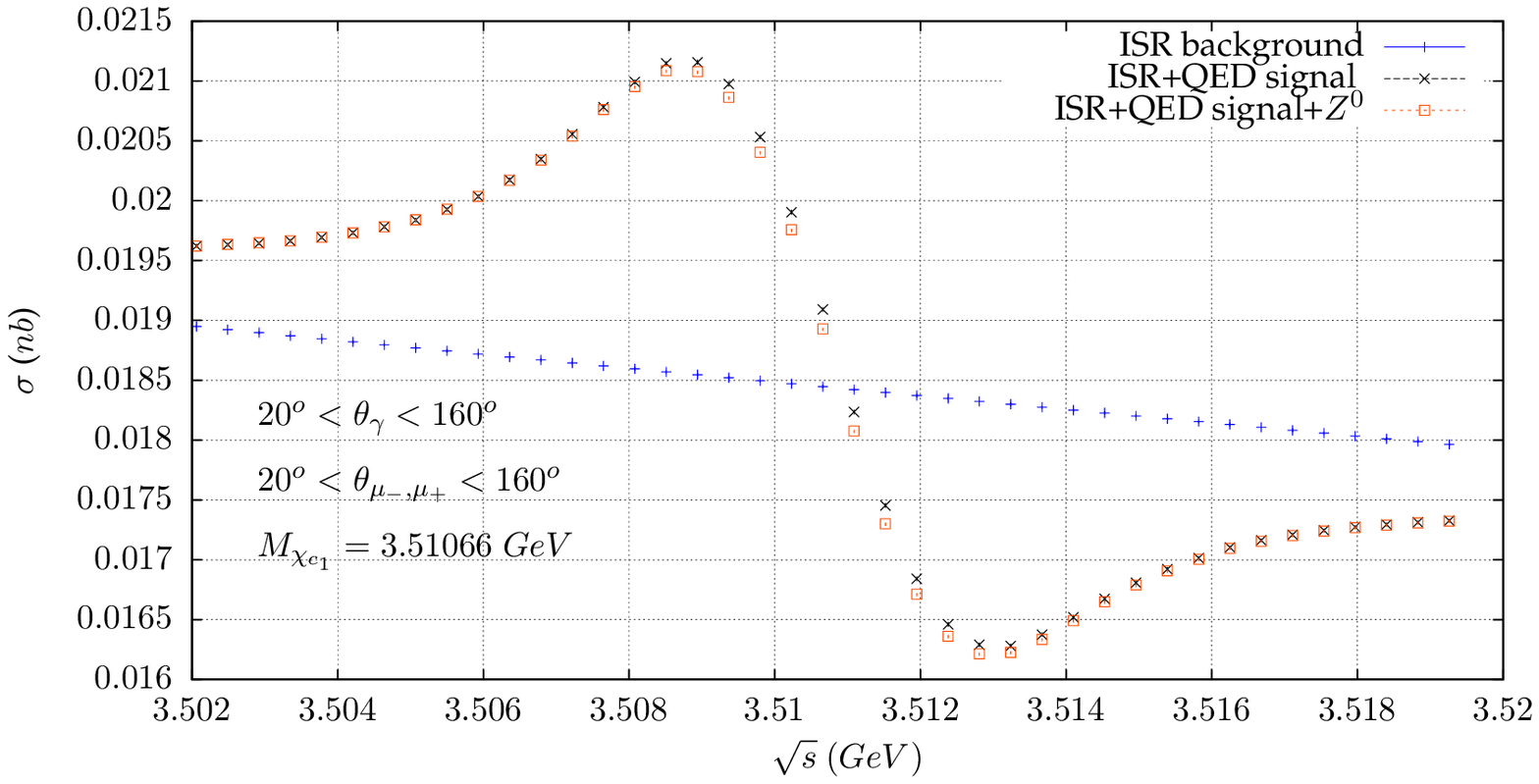}
\caption{The cross section $e^+ e^-\to \mu^+ \mu^- \gamma$, see text for details.
\label{cscutschi1}
}
\end{center}
\end{figure}

\begin{figure}
\begin{center}
\includegraphics[width=9.cm]{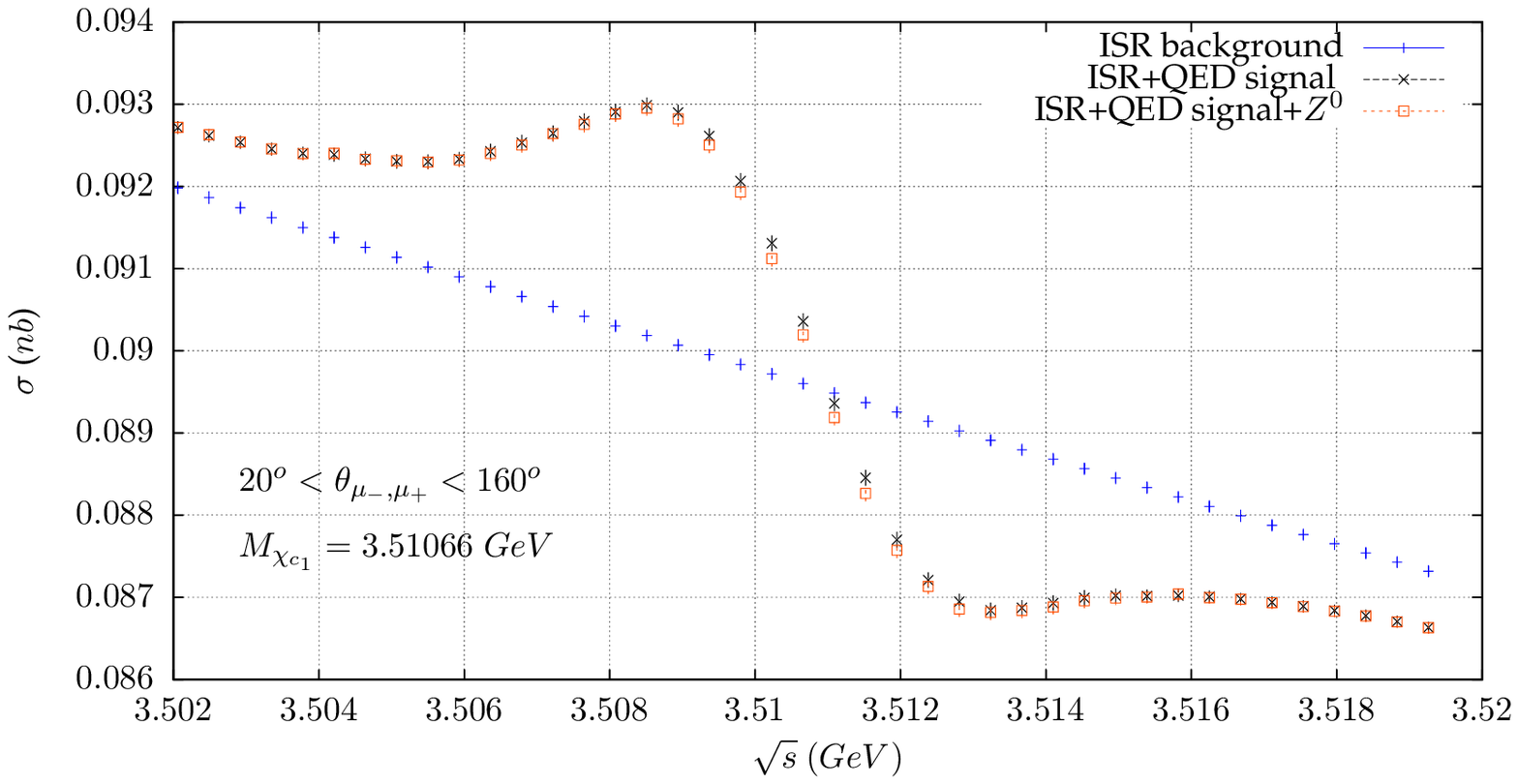}
\caption{The cross section $e^+ e^-\to \mu^+ \mu^- \gamma$, see text for details.
\label{cschi1}
}
\end{center}
\end{figure}
\begin{figure}
\begin{center}
\includegraphics[width=9.cm]{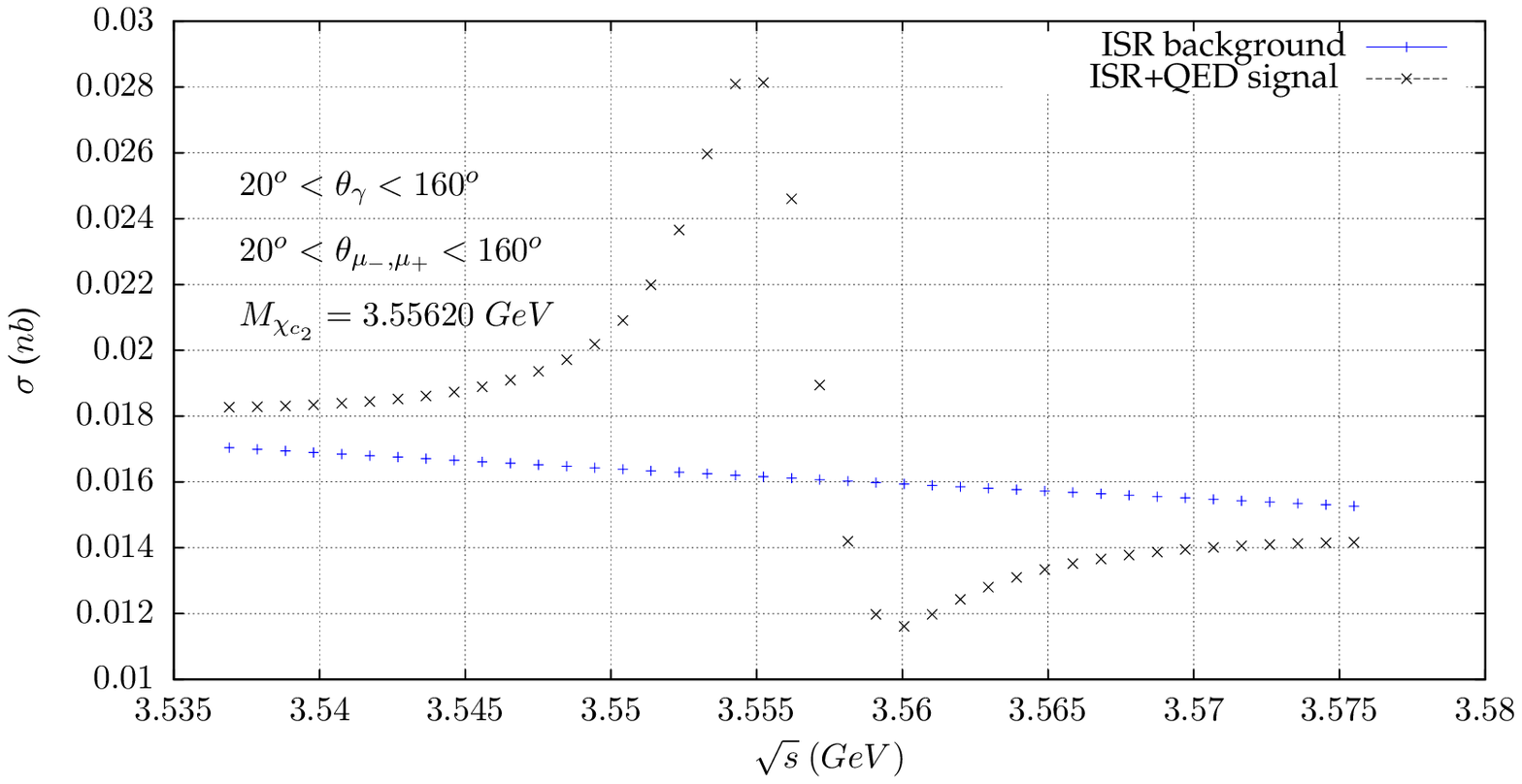}
\caption{The cross section $e^+ e^-\to \mu^+ \mu^- \gamma$, see text for details.
\label{cscutschi2}
}
\end{center}
\end{figure}
\begin{figure}
\begin{center}
\includegraphics[width=9.cm]{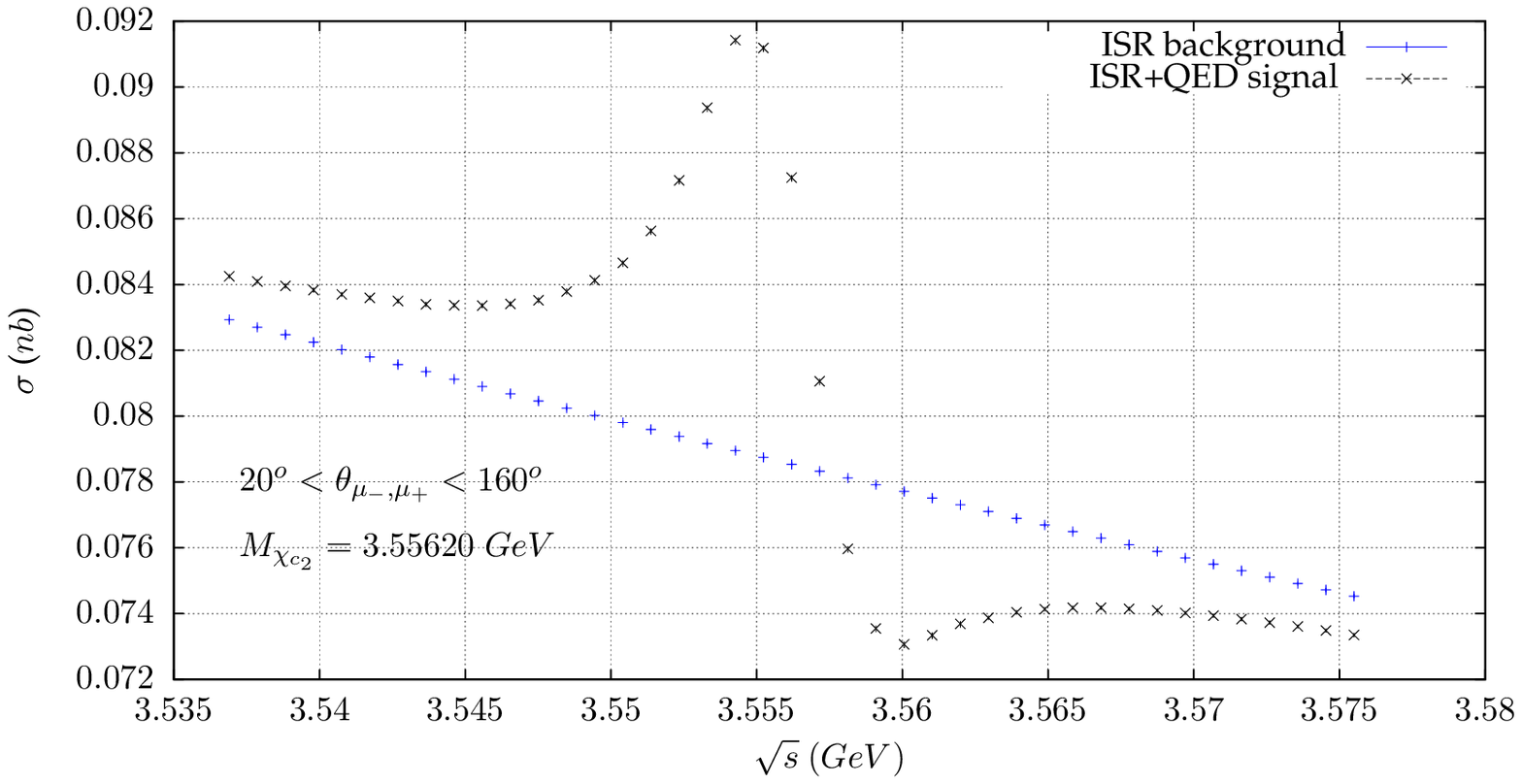}
\caption{The cross section $e^+ e^-\to \mu^+ \mu^- \gamma$, see text for details. 
\label{cschi2}
}
\end{center}
\end{figure}

The predicted values of the electronic widths are big enough to be observed
  at BESIII experiment \cite{Denig:2014fha}
  with scan at the vicinity of the $\chi_{c1}$ and 
  $\chi_{c2}$ at difference with other models 
  \cite{Kivel:2015iea,Denig:2014fha,Yang:2012gk}.  For $\chi_{c1}$ the 
  prediction of the electronic width is in agreement with the one 
    obtained within the 
  vector dominance model of \cite{Kuhn:1979bb}. 
  In Figures  \ref{cscutschi1} and
 \ref{cscutschi2} we show the cross sections of the reactions
  $e^+e^-\to \mu^+\mu^-\gamma$ imposing angular cuts on photons, whereas in Figures \ref{cschi1} and
 \ref{cschi2} we present these cross sections without imposing this cuts. In both cases we have assumed the $\chi_{c_1}$ and $\chi_{c_2}$
  electronic widths as listed in Table \ref{mod4_ee}. A beam spread of 
  $1\ MeV$ per beam with Gaussian distribution was assumed.
   Possible contributions
  from the diagrams in  Fig. \ref{cross}(b) and Fig. \ref{cross}(a) 
  with $J/\psi$ substituted with $\psi'$
  are negligible for event selections used in the plots, where 
  the muon pair invariant mass was chosen to be within 3 $J/\psi$ widths
  within $J/\psi$ mass (detector resolution was not taken into account).
  In the distributed version of the generator the diagrams with
  $\chi_{c_i} \to \gamma^*(\to \mu^+\mu^-) \gamma$ as well as
  $\chi_{c_i} \to \psi'^*(\to \mu^+\mu^-) \gamma$ are included. 
 As the contribution of $Z^0$ to the $\chi_{c_i}$ width is tiny, the same is expected for the diagram similar to Fig.\ref{cross}(c) with $\gamma$ substituted with $Z^0$ and these contributions were neglected.

 A signal of up to 75 \% of the radiative return
   background can be observed. The cross section is obviously 
  bigger, when the photon is not tagged, but the signal to background
  ratio is smaller and the BES-III collaboration will be able
  to measure these cross sections and extract the electronic widths 
  of the $ \chi_{c_1}$ and $ \chi_{c_2}$ if the model we present is correct.
  The scan in the vicinity
  of these two charm states would also provide the 
  possibility of testing the models and extracting the phase 
 between the radiative return and the $ \chi_{c_1}$ ($ \chi_{c_2}$) production
 amplitudes. As one can observe, with the relative phases between the amplitudes predicted within the model adopted in this paper, the production of $ \chi_{c_1}$ and $ \chi_{c_2}$ can be mainly observed as an interference between the ISR and the signal diagrams.

\section{\label{sec4} Conclusions}
Diract, resonant production of $\chi_{c1}$ and $\chi_{c2}$ in
 electron-positron annihilation through two virtual photons will lead to a
  measurable resonant enhancement of the cross
 section. The prediction exhibits a sizeable model dependence, a
 consequence of the fact that predictions for charmonium, based on the nonrelativistic
 potential model are of qualitative nature only. Nevertheless, a
 resonant signal both in the hadronic cross section and in the
 $\gamma\mu^+\mu^-$ channel could be seen at the BESIII storage ring under
 favorable circumstances.

\begin{acknowledgments}
We would like to thank A. Denig for discussions of the experimental aspects of our analysis.
\end{acknowledgments}
\appendix

\section{$g_1$ and $g_2$ couplings}
\label{g1g2}
The  effective couplings $g_1$ and $g_2$ 
  are defined in Section \ref{sec2} through loop integrals. 
 We split them into two parts. One coming from diagrams
  containing $\chi_{c_i}-\gamma-\gamma$ vertex and called 
  $g_{i_{\gamma \gamma}}$, diagrams containing $\chi_{c_i}-J/\psi-\gamma$
  vertex called $g_{i_{J/ \psi \gamma}}$ and diagrams containing 
  $\chi_{c_i}-\psi'-\gamma$
  vertex called $g_{i_{\psi' \gamma}}$. The constants $g_1$ and $g_2$ 
  are sums of these
 three contributions
  $g_i=g_{i_{\gamma \gamma}}+g_{i_{J/ \psi \gamma}}+g_{i_{\psi' \gamma}}$.

The couplings read ($M=M_{\chi_{c_1}}$ in $g_1$;
 $M=M_{\chi_{c_2}}$ in $g_2$; $M_J\equiv M_{J/\psi} $ in $g_{i_{J/ \psi \gamma}}$);
  $x\equiv 4m^2/M^2$, $y\equiv 4M^2_{J/\psi}/M^2$
 
\bea
g_{1_{\gamma \gamma}}&=&  \frac{16\alpha^2a}{\sqrt{m}M^2}\Bigg[ 
  \log\left(\frac{x}{1+x}\right)
   \left(1-x\right)\nonumber\\
  &&-\left(\log\left(\frac{x}{1-x}\right)+i\pi\right)
   \left(1+x\right)\Bigg] ,
\eea

\bea
  g_{2_{\gamma \gamma}}&=&   \frac{32\sqrt{2}\alpha^2a}{3\sqrt{m}M^2} 
 \Bigg[\left( \frac{1+x}{2}+\frac{8}{(1+x)^2}\right)\log(1-x)\nonumber\\
 &&\kern-32pt+ \frac{3}{2}\left(1+x\right) \log(1+x)
 -2\left(1+x + \frac{2}{(1+x)^2}\right)\log(x)\nonumber\\
 &&\kern-32pt-\frac{8}{(1+x)^2}\log(2) -1 
  -\frac{i\pi}{2}\left( 1+x +\frac{8}{(1+x)^2}\right) 
\Bigg]\nonumber\\
 \eea

\bea
g_{1_{J/ \psi \gamma}}
&=& \frac{8\alpha^2a_J f}{\sqrt{4\pi\alpha m}M^2M_J^2}\Bigg[ 
 \left(\log\left(\frac{x}{1-x}\right)+i\pi\right)
  \left(1+x-\frac{y}{2}\right)\nonumber\\
 && +F_0(x,y)-\frac{1}{4}\left(3+x+y \right)F_1(x,y)
 \nonumber\\
 && -\frac{y(4+y)}{2(2+2x-y)^2} F_2(x,y)
  +\frac{y(1+y-x)}{2(2+2x-y)}F_3(x,y)\nonumber\\
 &&- \frac{y}{2}F_4(x,y)+\frac{y}{2}\left(3-x\right)F_5(x,y)\Bigg],\nonumber\\
\eea

\bea
g_{2_{J/ \psi \gamma}}&&=\frac{16\sqrt{2}\alpha^2 a_J f}{3\sqrt{4\pi\alpha m}M^2M_J^2}
 \Bigg[ 2 -\log(2)\left(3-\frac{16}{(1+x)^2}\right) \nonumber\\ 
 && +\log(x)\left(1-y + 2x + \frac{8} {(1+x)^2} \right) \nonumber\\
  &&+\log(1-x)\left(\frac{1}{2} +y-2x-\frac{16}{(1+x)^2}\right)\nonumber\\
  &&-\frac{3y}{8}\log\left(\frac{y}{4}\right)
   +\log\left(1-\frac{y}{4}\right)
   \left(-\frac{3}{2}+\frac{3y}{8}\right)\nonumber\\
  &&+i\pi\left( 1-\frac{11y}{8}+2x+ \frac{8} {(1+x)^2}\right)\nonumber\\
 &&-F_0(x,y)-\left(\frac{1}{2}+y-\frac{x}{4}\right)F_1(x,y)\nonumber\\
  &&\kern-32pt+\frac{-55-123xy+126x+93x^2-94y+38y^2}{16(2+2x-y)^2}F_2(x,y)\nonumber\\
   &&+\frac{87-5xy-2y+2y^2+2x+3x^2}{2(2+2x-y)}F_3(x,y)\nonumber\\
   &&-\frac{3y}{4}F_4(x,y)-\frac{3y}{4}(1+x)F_5(x,y)\Bigg]
\eea

with

\bea
  r&=&\sqrt{x- (1-y+x)^2/4}
\eea
 and
\bea
 &&\kern-30pt A(x,y) = \arctan\left(\frac{1-y+x}{2r}\right)
  -\arctan\left(\frac{-1-y+x}{2r}\right)\nonumber\\
  && F_0(x,y) =  \frac{1+y-x}{4}\log(x/y)-r A(x,y)\nonumber\\
 && F_1(x,y) = \log(x/y)+ \frac{1+y-x}{r} A(x,y)\nonumber\\
 && F_2(x,y) = 2\log(2)-x\log(x)+y/2\log(y/2)\nonumber\\
  &&-(1-x)\left(\log(1-x)-i\pi\right)
  \nonumber\\
 && +(2-y/2)\left(\log(2-y/2)-i\pi\right)\nonumber\\
 &&+\frac{-1-x+y}{2}\log(x)+\frac{-1+x-y}{2}\log(y)\nonumber\\
 &&-2rA(x,y)\nonumber\\
 &&F_3(x,y) =  -\frac{3}{2}\log(x)+ \log(1-x)-i\pi
   \nonumber\\
 && +\frac{1}{2}\log(y)
   -\frac{1-x+y}{2r}A(x,y)\nonumber\\
  && F_4(x,y) = \log(1-2/y)\log(y/2) -\Li2(2/y)\nonumber\\
  && +\Li2\left(\frac{1-y/2}{1+x-y/2}\right)-\Li2\left(\frac{-y/2}{1+x-y/2}\right)\nonumber\\
 &&- \Li2\left(\frac{1-y/2}{(1-x)/2+ir_1}\right)
   - \Li2\left(\frac{1-y/2}{(1-x)/2-ir_1}\right) \nonumber\\
  &&+ \Li2\left(\frac{-y/2}{(1-x)/2+ir_1}\right)
   + \Li2\left(\frac{-y/2}{(1-x)/2-ir_1}\right)\nonumber\\
  && F_5(x,y) =  -\frac{1}{1+x-y/2}
  \log\left(\frac{1+x}{x}\right)\nonumber\\
  &&+\frac{-r_1+i(1+y-x)/2}{(1-x+2ir_1)r_1}
   \log\left(\frac{(1-x+y)/2+ir_1}{(-1-x+y)/2+ir_1}\right)\nonumber\\
  &&-\frac{r_1+i(1+y-x)/2}{(1-x-2ir_1)r_1}
   \log\left(\frac{(1-x+y)/2-ir_1}{(-1-x+y)/2-ir_1}\right)\nonumber\\
\eea
with
\bea
  r_1&=&\sqrt{x- (1+y-x)^2/4} .
\eea

 For the $g_{i_{\psi' \gamma}}$ the expressions are similar to $g_{i_{J/\psi \gamma}}$
 with the following changes: $M_J\equiv M_{\psi'} $, $y\equiv 4M^2_{\psi'}/M^2$,

 \bea
  i\cdot r_1&=& - \sqrt{ (1+y-x)^2/4-x} ,
\eea
 \bea
 \log(2-y/2) \to \log(y/2-2) + i\pi
\eea
 \bea
 \log(1-y/4) \to \log(y/4-1) + i\pi
\eea
\bea
 \kern-25pt\frac{A(x,y)}{r} &=& \frac{1}{2\tilde r}
\Biggl\{ \log\left[\frac{(1-y+x)/2 - \tilde r}{(1-y+x)/2 + \tilde r}\right]
      \nonumber \\
      &&\kern+15pt- \log\left[\frac{(-1-y+x)/2 - \tilde r}{(-1-y+x)/2 + \tilde r}\right]
  \Biggr\}
\eea
 
with 
\bea
 \tilde r = \sqrt{(1-y+x)^2/4-x}.
\eea

\bibliography{biblio}

\end{document}